AJP Resource Letter

# Gravitational Lensing

*Tommaso Treu,*[1] *Philip J. Marshall,*[2] *Douglas Clowe* [3]

1. Dept. of Physics,, University of California Santa Barbara, CA 93106, USA
2. Dept. of Physics, University of Oxford, Keble Road, Oxford, OX1 3RH, UK
3. Dept. of Physics & Astronomy, Ohio University, Clippinger Lab 251B, Athens, OH 45701, USA

This Resource Letter provides a guide to a selection of the literature on gravitational lensing and its applications. Journal articles, books, popular articles, and websites are cited for the following topics: foundations of gravitational lensing, foundations of cosmology, history of gravitational lensing, strong lensing, weak lensing, and microlensing.

## I. INTRODUCTION

Gravitational lensing is the astrophysical phenomenon whereby the propagation of light is affected by the distribution of mass in the universe. As photons travel across the universe, their trajectories are perturbed by the gravitational effects of mass concentrations with respect to those they would have followed in a perfectly homogeneous universe. The description of this phenomenon is similar in many ways to the description of the propagation of light through any other media, hence the name gravitational lensing or gravitational optics.

It is customary to distinguish three lensing regimes: strong, weak, and micro. Strong lensing is said to occur when multiple images of the source appear to the observer; this requires a perturber creating a strong gravitational field, and very close alignment of the lens and source. In general, the gravitational field of the deflector is not strong enough to create multiple images, and the observable effect is just a generic distortion of the images detectable only in a statistical sense; this is called weak lensing. For historical and practical reasons, strong lensing producing very small angular separations between the multiple images is called microlensing; the name comes from the typical angular separation created by the gravitational field of a star, which is typically of order microarcseconds.

Gravitational lensing is a relatively young field. Strong lensing, weak lensing, and microlensing observations are only a few decades old. Yet, in just a short time it has gone from a curiosity, mostly appreciated for its aesthetic and mathematical appeal, to a powerful tool used to study an impressive range of astrophysical phenomena, from planets to galaxies, clusters of galaxies, dark matter and dark energy.

Our aim is to provide an entry point into the extraordinarily rich and diverse literature on gravitational lensing, from its theoretical foundations to its astrophysical applications. Given the breadth of the subject, it is impossible to be complete. By necessity, our choices will exclude many important resources and perhaps even entire topics and subtopics. We apologize in advance to those authors whose contributions to the field we have overlooked. The resources and commentary we cite should be sufficient to locate useful resources in addition to the ones presented here. In this spirit, when possible we have chosen review and modern articles over older work. We have also included some resources on general cosmology, necessary to make sense of some of the details of weak and strong gravitational lensing in the universe, and some of their applications. Although general relativity is at the heart of gravitational lensing theory, we do not refer to any general relativity resources here, unless they are specifically devoted to gravitational lensing. Useful resources on general relativity can be found, for example, in

1. "Resource Letter BH-2: Black holes," E. Gallo, D. Marolf, Am. J. Phys. **77**, 294-307, (2009). (E)

This Resource Letter is organized as follows. In the remainder of this section we introduce some basic resources referred to throughout. In Section II we present popular material, including some covering the history of gravitational lensing. In Section III we present material suitable for undergraduates, primarily textbooks. In Section IV we list graduate-level textbooks and general reviews. Specialized material from the research literature is presented in Sections V-VII, sorted by topic as far as possible. We begin with strong gravitational lensing in Section V, continue with weak gravitational lensing in Section VI, and conclude with microlensing in Section VII.

*Basic resources:*

Online resources, journals, and reviews provide a natural entry point into a new subject. Many useful resources will appear after the publication of this Resource Letter; what follows is a non-exhaustive list of websites and publications that are currently useful and are likely to continue to be useful in the future.

2. *The SAO/NASA Astrophysics Data Systems (ADS) Abstract Service,* adsabs.harvard.edu/abstract_service.html This website is the standard search engine for astrophysics-related publications of all sorts. It allows all kinds of searches, including by authors, journal and publication date, and by specific objects, tools, and title or abstract keywords. It also links and keeps track of references, to allow searches by citation or reference, and to enable rough measures of impact. The database contains scholarly research papers, as well as conference proceedings, proposal abstracts, and books. (A)
3. *Annual Reviews of Astronomy and Astrophysics,* http://arjournals.annualreviews.org/loi/astro This journal publishes reviews written by well-known proponents of their specialized fields, at the level of a beginning graduate student in physics or astrophysics. (A)
4. *Astrophysical Journal,* http://iopscience.iop.org/0004-637X One of the leading American journals publishing specialized astrophysics articles. It is "devoted to recent developments, discoveries, and theories in astronomy and astrophysics." Short papers describing very new results are published as "letters," abbreviated as Astrophys. J.L. (A)
5. *Astronomical Journal,* http://iopscience.iop.org/1538-3881 Sister journal to the Astrophys. J., the Astron. J. "publishes original astronomical research, with an emphasis on significant scientific results derived from observations, including descriptions of data capture, surveys, analysis techniques, and astronomical interpretation." (A)

6. *Monthly Notices of the Royal Astronomical Society,* http://www.wiley.com/bw/journal.asp?ref=0035-8711  This long-standing British journal "publishes the results of original research in positional and dynamical astronomy, astrophysics, radio astronomy, cosmology, space research and the design of astronomical instruments." (A)
7. *Astronomy & Astrophysics,* http://www.aanda.org   One of the leading European astrophysics journals, *Astron. Astrophys*. "publishes papers on all aspects of astronomy and astrophysics (theoretical, observational, and instrumental) independently of the techniques used to obtain the results." (A)
8. *Living Reviews in Relativity***,** http://relativity.livingreviews.org   Open-access web-based peer-reviewed journal, publishing specialized reviews in all areas of relativity, including gravitational lensing. (A)
9. *New Journal of Physics,* http://www.njp.org   This open-access journal "publishes across the whole of physics, encompassing pure, applied, theoretical and experimental research, as well as interdisciplinary topics where physics forms the central theme." (A)
10. *Advances in Astronomy,* http://www.hindawi.com/journals/aa   This is an online publication with good reviews in all topics of astrononomy. (A)
11. The e-print archive, or "astro-ph", http://arxiv.org   The open-access arxiv is an online electronic archive of user-submitted articles in a range of mathematical science subjects; astro-ph is its astrophysics wing. Most authors now submit their journal articles to astro-ph either at the time of their submission to a journal, or upon their acceptance. In the former case, astro-ph gives the community a view of a research paper, and a chance to comment on it, while it is going through the at times lengthy review process. One significant advantage of the arxiv is that it is open access: most journal websites must be subscribed to, at a cost significant enough that only libraries can afford to pay it.  The arxiv makes astronomical research papers available to anyone with an internet connection. Note that you can reach articles posted on astro-ph via the ADS (ref 2) interface. (A)
12. *Astronomy and Geophysics***,** http://www.wiley.com/bw/journal.asp?ref=1366-878  This is another publication of the Royal Astronomical Society. In addition to research articles, it publishes news and reviews accessible to intermediate-level audiences. (I)
13. *Scientific American,* http://www.scientificamerican.com   This popular general science magazine often includes excellent articles on astronomy and astrophysics accessible to the general public. (E, I)
14. *Nature,* http://www.nature.com   A research journal publishing articles in astrophysics, as well as in all other branches of science. It strives to be accessible to scientists outside its individual articles' fields. It also publishes news and views, i.e. comments on recent articles considered to be of high impact. (I, A)
15. *Science*, http://www.sciencemag.org   A research journal publishing articles in all areas of science, including astrophysics. It strives to be accessible to scientists outside their fields. (I, A)
16. *Astronomy Magazine***,** http://www.astronomy.com    A very popular magazine among amateur astronomers and astronomy buffs. It often includes beautiful pictures, as well as articles and tips for astrophiles. (E, I)

17. *Physics Today*, http://www.physicstoday.org  This magazine is published by the American Institute of Physics, and includes excellent popular articles on all areas of physics, including astrophysics. (E, I)

## II. POPULAR MATERIAL

This section is intended for persons with a high-school education, but without college-level specialized training in physics or astronomy.

*A. Lensing in the Popular Science Literature*

18. **In Search of Dark Matter,** K. Freeman and G. MacNamara, (Springer, Berlin, 2006). In this focused look at the dark-matter problem, gravitational lensing is again introduced and explained to the lay reader. (E, I)
19. **Einstein's Telescope,** Evalyn Gates (W.W. Norton & Company, New York, 2009). A popular book that describes how gravitational lensing can be used to answer fundamental questions like "What is the Universe made of?" (E,I)

*B. Lensing in the News*

Several scientific results based on gravitational lensing have made headline news in the last decade, with their stories seemingly very effective at capturing the public's imagination. Important themes seem to include "seeing" the invisible, be it planets, dark matter in clusters, or the large scale structure in the Universe, using lensing to understand the dark energy-driven expansion of the Universe, and constructing giant "cosmic telescopes" that make use of the natural magnification provided by gravitational lenses. Most news sources refer back to the press release source (where links to the journal articles can be found); this is especially true of web-based science magazines. Here we give pointers to two examples of particularly high-profile distributors running the stories.

20. "The Universe Gives Up Its Deepest Secret," S. Connor, *The Independent,* 8 January 2007. One of the results of the analysis of weak lensing in the COSMOS field was a three-dimensional mass map, described by Massey *et al* as "the scaffolding of the Universe." Many news outlets covered the COSMOS dark matter map story, but in *The Independent* the map was presented covering the whole of the front page. (E)
21. "Worlds Collide", *The New York Times,* 23 August 2006. COSMOS was by no means the first time that dark matter mapping with gravitational lensing had attracted publicity. The year before, the lensing analysis of the Bullet Cluster was picked up as providing the "proof" of dark matter's existence.

*C. Lensing Through History*

The history of the discovery of the gravitational lensing has some very interesting features. Most notably, it has seen some key theoretical predictions made, which were later confirmed by observation. This is somewhat atypical for astronomy, which tends to be observationally-driven, with theoretical explanations for new phenomena following behind. Several of the more accessible, general reviews cited in the sections below contain useful historical notes: in particular, see references 48, 54, 86, 87.

As one of the first observational tests of Einstein's general relativity, the famous 1919 eclipse expedition is an excellent story:

22. "Gravitational lensing: a unique probe of dark matter and dark energy," R. S. Ellis, Phil. Trans. R. Soc.,**368**, 967-987, (2010). An excellent overview of the history of gravitational lensing, including the 1919 eclipse, covering the history of the subject to the present day. (I,A)

From a historical perspective, three articles in particular about the eclipse expedition are worth highlighting:

23. "A Determination of the Deflection of Light by the Sun's Gravitational Field, from Observations made at the Total Eclipse of May 29, 1919," F. W. Dyson, A. S. Eddington, and C. R. Davidson, Phil. Trans. R. Soc. A. **220**, 291-333, (1920). The original paper describing the expedition, its observations and results. (A)
24. "Testing relativity from the 1919 eclipse - a question of bias," D. Kennefick, Physics Today **62**, 37-42, (March 2009). http://dx.doi.org/10.1063/1.3099578 tells the story of Dyson and Eddington's experiment and data analysis. Eddington believed that the experiment would confirm Einstein's prediction - and his observations bore this out. (I)
25. "An Expedition to Heal the Wounds of War," M. Stanley, Isis **94,** 57-89, (2003). Eddington, a pacifist, used the eclipse experiment as an example of how science can transcend borders, even in times of war. (I)

The deflection of starlight during an eclipse was the first observation of lensing in the universe, and there would not be another for 60 years. However, in the 1930s, Albert Einstein considered lensing by stars other than the sun (what we now call microlensing), and, following Henry Norris Russell's ensuing 1937 Scientific American article "A Relativistic Eclipse," Fritz Zwicky predicted that galaxies and clusters of galaxies would make good gravitational lenses as well.

26. "Lens-like action of a star by the deviation of light in the gravitational field," A. Einstein Science **84** (2188), 506–507 (1936). This classic paper marks the birth of microlensing, in making the jump from light deflection by our sun (as observed by Eddington, et al.) to that by other stars in our galaxy. (A)
27. "Nebulae as gravitational lenses," F. Zwicky, Phys. Rev. **51**, 290–290 (1937). http://dx.doi.org/10.1103/PhysRev.51.290 (A)

The sources behind such lenses would be galaxies and their active nuclei, in the far-distant universe; only in the last 30 years have such faint objects been detected in sufficient numbers to reveal the one in a thousand that are multiply-imaged. The first lensed quasar was confirmed.

28. "0957+561 A, B - Twin quasistellar objects or gravitational lens," D. Walsh, R. F. Carswell, and R. J. Weymann, Nature **279**, 381–384 (1979). http://dx.doi.org/10.1038/279381a0 (A)

Quiescent galaxies are more numerous sources, but are even fainter than quasars: it would take the advent of CCD imaging cameras to detect the first gravitational arc.

29. "The giant arc in A 370 - Spectroscopic evidence for gravitational lensing from a source at Z = 0.724," G. Soucail, Y. Mellier, B. Fort, G. Mathez, and M. Cailloux, Astron. Astrophys. **191**, L19-L21 (1988). It was not clear what the arc was: a deep spectrum revealed it to be a background galaxy, observed at high magnification through the lens. (A)

**III. UNDERGRADUATE MATERIAL**

*A. Textbooks*

Since gravitational lensing is just one astronomical phenomenon, and one observational consequence of General Relativity, it does not have entire undergraduate textbooks devoted to it. However, it does feature in each of the following books:
30. **Introduction to Cosmology**, B. S. Ryden (Addison Wesley, San Francisco, 2003). An excellent introduction to cosmology and starting point to the study of gravitational lensing. Suitable for undergraduates with basic knowledge of calculus. (I,A)
31. **Cosmological Physics,** J. A. Peacock (Cambridge University Press, Cambridge, 2002). A classic cosmology textbook, with a chapter on gravitational lensing. Suitable for advanced undergraduates and graduate students. (A)
32. **Extragalactic Astronomy and Cosmology**, P. Schneider (Springer-Verlag, Heidelberg, 2006). A textbook aimed at advanced undergraduate and beginning graduate students that incorporates derivations of gravitational lensing in a cosmological and galaxy-evolution framework. (I,A)
33. **Gravity from the Ground Up,** B. Schultz (Cambridge University Press, Cambridge, 2003). Containing a chapter on gravitational lensing, this textbook is aimed at general readers and undergraduates, and assumes only high-school level mathematics. (E, I)

*B. Articles*

Gravitational lensing has been covered a number of times by popular science magazines, at an undergraduate level. Here is a representative selection of articles.
34. "Gravitational Lenses," E. L. Turner, Sci. Am., **259,** 26-32, (July 1988). (I)
35. "Gravity's Kaleidoscope," J. Wambsganss, Sci. Am., **285**, 52-59 (November 2001). (I)

While not available online, the above two articles give introductions to strong gravitational lensing in general and microlensing in particular.
36. "Gravitational Lenses," L. V. E. Koopmans and R. D. Blandford, Physics Today, **57**, 45-51 (June 2004). This overview is available online, and also focuses on strong lensing. (I)
37. "Stars that Magnify Quasars," S. Refsdal and R. Stabell, New Scientist, **123,** 51-53 (July 1989). An exposition of the phenomenon and uses of quasar microlensing. (I)

In the weak lensing regime, feature-length articles are rarer. A notable exception is:
38. "Mapping Dark Matter with Gravitational Lenses," J. A. Tyson, Physics Today, **45**, 24-32, (June 1992). Introduces and explains the process of inferring a cluster of galaxies' mass distribution, from weak shear data. (I, A)

These publications also provide online resources.

39. "Dark Matter," Sci. Am., (November 2010) multimedia interactive feature. http://www.scientificamerican.com/media/DarkMatter/landing.html

*C. Practical Activities*

Gravitational lensing can be demonstrated and explored using physical models with equivalent optical properties - the base of a wineglass is an excellent example. Optical gravitational-lensing experiments make an interesting astrophysics-themed addition to an undergraduate teaching laboratory program:

40. "The Optical Gravitational Lens Experiment," J. Surdej, S. Refsdal, and A. Pospieszalska-Surdej, in **Gravitational Lenses in the Universe, Proceedings of the 31st Liege International Astrophysical Colloquium (LIAC 93), held June 21-25, 1993**, edited by J. Surdej, D. Fraipont-Caro, E. Gosset, S. Refsdal, and M. Remy, (Institut d'Astrophysique, Liege, 1993), pp.199-203. (I) See also the website at http://wela.astro.ulg.ac.be/themes/extragal/gravlens/bibdat/engl/DE/didac.html

This article, and the associated webpages, were purposely written for educators: the text, diagrams and photographs are clear and didactical.

*D. Software*

The most suitable undergraduate-level software available takes the form of code running on interactive web pages. Simple downloadable applications are also available.

41. "Ned Wright's Cosmology Tutorial," E. L.Wright, http://www.astro.ucla.edu/~wright/cosmolog.htm An excellent entry point into cosmology, with plenty of very interesting and effective webpages, comments on scientific news and a java cosmological quantity calculator. (I, A)
42. "Mowgli," P. S. Naudus, P. J. Marshall and J. F. Wallin, http://ephysics.org/mowgli This Java applet allows the user to explore simple models for the lens mass and source light, by manipulating the model components on the screen with the mouse. (I,A)
43. "HST Gravitational Lens Image Creator," Medium Deep Survey team. http://virtual-universe.org/ego_cgi.html With the emphasis on generating mock data rather than modeling real data, this Javascript tool takes input lens, source and HST WFPC2 observation parameters and generates a predicted JPG image. (I,A)

**IV. ADVANCED MATERIAL: GENERAL**

In the remainder of this Resource Letter we suggest a number of resources from textbooks to articles and websites on gravitational lensing and its astrophysical applications for advanced readers. Resources of general interest are listed in this section, while Sections V, VI, VII specialize in strong, weak, and microlensing material. Some of the general resources include sections on each of these subtopics. After listing reviews and textbooks, we list for each subtopic a few examples of research papers. We have tried as much as possible to list recent papers to capture and summarize the state of the art of a field and, when possible, seminal papers that shaped the field. Our list is by no means complete. We refer the reader to the reviews and introductions of the recent papers for a more comprehensive list of references to explore.

*A. Textbooks and Lectures on gravitational lensing and cosmology*

44. **Gravitational Lensing: Strong, Weak, and Micro,** P. Schneider, C.S. Kochanek, and J. Wambsganss (Springer-Verlag, Berlin, Heidelberg, 2006) [Saas Fee Advanced Course 33], edited by G. Meylan, P. Jetzer and P. North. A comprehensive and pedagogical set of lectures, providing an ideal first step for graduate students and researchers entering the field. Includes a very useful introduction to cosmology. (A)
45. **Gravitational Lenses,** P. Schneider, J. Ehlers, and E.E. Falco (Springer, Berlin, 1992). The classic monograph on gravitational lensing. A must read for advanced readers. (A)
46. "Cosmology with Gravitational Lensing", A. Heavens, in **Dark Matter and Dark Energy** [ASSL 370], edited by S.Matarrese, M.Colpi, V.Gorini & U.Moschella (Springer, Berlin, 2011). These lectures were given at the Como Summer School 2007, and include a simple worked derivation of the lens equation in general relativity. [arXiv:1109.1121](arXiv:1109.1121) (A)
47. **Singularity Theory and Gravitational Lensing**, A.O. Petters, H. Levine, and J. Wambsganss (Birkhauser, Boston, 2001). This is the only monograph to focus on the pure mathematical aspects of gravitational lensing. An excellent book suitable for advanced readers with a strong interest in mathematics. (A)

*B. Reviews*

48. "Gravitational lensing," M. Bartelmann, Classical and Quantum Gravity **27** (23), 233001, 72-156 (2010). An up-to-date and comprehensive review of all aspects of gravitational lensing, from theory to observations. Includes derivations from first principles of gravitational lensing relations. (A)
49. "Gravitational Lensing in Astronomy," J. Wambsganss, Living Rev. Relativ., **1**, 12 (1998), http://www.livingreviews.org/lrr-1998-12 (cited on 01/16/2012). A comprehensive (and updatable) review with sections on the history, theory, and applications of gravitational lensing. (A)
50. Large Synoptic Survey Telescope (LSST) Science Book, Chapters 12, 14, and 15, LSST Collaboration, arXiv:0912.0201 (2009). This book was written to describe the science that is anticipated being performed using the Large Synoptic Survey Telescope, a future optical imaging facility. It provides an excellent starting point for exploring the current plans for making gravitational lensing measurements from future large area surveys, and what those measurements will enable. (A)
51. "The dark matter of gravitational lensing" R. Massey, T. Kitching, and J.Richard, Rep. Prog. Phys. **73**, 086901 (2010). This recent review is focused mostly on clusters and large-scale structure. (A)
52. "Cluster Lenses" J.-P. Kneib, P. Natarajan, Astron Astrophys Rev. **19**, 47-146 (2011). This recent comprehensive review covers strong and weak lensing theory and observations, focusing on galaxy clusters as deflectors. (A)

**V. ADVANCED MATERIAL: STRONG LENSING**

Strong gravitational lensing has many applications in solving important astrophysical problems, primarily for three reasons: it can be used to measure precisely the gravitational potential of astronomical objects, to measure distances, and to magnify distant objects. We give here a few examples, including the study of dark matter in galaxies and clusters of galaxies, the study of the geometry and content of the universe, and the study of the properties of the most distant galaxies and quasars. Strong lensing is a highly technical field: one needs sophisticated algorithms and statistical methods to find lenses and model them to extract useful information. For this reason we conclude this section by listing resources for some of these more technical aspects. In the modeling section we give links to publicly available software that we consider state of the art.

*A. General strong lensing material*

53. "Strong lensing by galaxies," T. Treu, Ann. Rev. Astron. Astrophys., **48**, 87-125 (2010). A review of the many astrophysical applications of strong gravitational lensing, focused on galaxy-scale deflectors. It includes applications to the study of the properties of galaxies, cosmography, and the use of lenses as natural telescopes. (A)
54. "A most useful manifestation of relativity: gravitational lenses," E.E. Falco, N.J.Phys. **7**, 200 [25 pages] (2005). A mainly observational review of the astrophysical and cosmological applications of strong gravitational lensing, including a nice historical introduction and a brief review of microlensing. (A).
55. "Quasar Lensing," F. Courbin, P. Saha, and P. Schechter, in **Gravitational Lensing: An Astrophysical Tool**, edited by F. Courbin, and D. Minniti, Lect. Not. Phys. **608**, 1-54 (2002). A detailed review of strongly lensed quasars, with dual emphasis on gravitational time delays as a tool for cosmology, and on cosmological microlensing. (A)
56. "Cosmological applications of gravitational lensing," R.D. Blandford and R. Narajan, Ann. Rev. Astron. Astrophys. **30**, 311-358 (1992). A classic review of the applications of strong gravitational lensing at a time when all strong lenses known could be discussed one by one. Still a very useful reference for the basic concepts. (A)
57. Online lens databases. CASTLeS, http://www.cfa.harvard.edu/castles This is a database of galaxy-scale gravitational lenses, listing their coordinates, their image astrometry, photometry and spectroscopic data and, for some systems, a list of references. The database is not complete, and some of the data are unreliable or out of date, so it is important to double check the information (A). The Master Lens Database http://masterlens.astro.utah.edu is the intended successor to CASTLeS: it contains much more extensive information (including lens model parameters) for a much larger, almost complete, sample of lenses. (A)

*B. Dark matter and stars in galaxies and clusters*

58. "Massive dark matter halos and evolution of early-type galaxies to z~1," T. Treu and L.V.E. Koopmans, Astrophys. J. **611**, 739–760, (2004). A clear illustration of the power of combining gravitational lensing with other diagnostics (in this case stellar kinematics) for the study of stars and dark matter in galaxies and their evolution over cosmic time. (A)
59. "The Sloan Lens ACS Survey. X. Stellar, Dynamical, and Total Mass Correlations of Massive Early-type Galaxies," M. W. Auger, T. Treu, A. S. Bolton, R. Gavazzi, L. V. E.

Koopmans, P. J. Marshall, L. A. Moustakas, and S. Burles, Astrophys. J. **724**, 511-525 (2010). A recent and up-to-date paper on the distribution of stellar and dark matter in early-type galaxies using strong lensing, stellar kinematics, and other astrophysical tools. (A)

60. "The Dark Matter Distribution in A383: Evidence for a Shallow Density Cusp from Improved Lensing, Stellar Kinematic, and X-ray Data," A. Newman, T. Treu, R. S. Ellis, and D. J. Sand, Astrophys. J. **728**(2), L39-L45 (2011). A good example of astronomers attempting to combine information from several sources to constrain a self-consistent model of a cluster gravitational potential. (A)
61. "The SWELLS survey - I. A large spectroscopically selected sample of edge-on late-type lens galaxies," T. Treu, A. A. Dutton, M. W. Auger, P. J. Marshall, A. S. Bolton, B. J. Brewer, D. C. Koo, and L. V. E. Koopmans, Mon. Not. R. Astron. Soc. **417**, 1601-1620 (2011). A recent description of the rationale for studying spiral galaxies with strong gravitational lensing and the challenges associated with finding large samples of them. (A)
62. "Dark Matter Contraction and the Stellar Content of Massive Early-type Galaxies: Disfavoring Light Initial Mass Functions," M. W. Auger, T. Treu, R. Gavazzi, A. S. Bolton, L. V. E. Koopmans, and P. J. Marshall, Astrophys. J. **721**(2), L163-L167 (2010). A concise paper that combines strong and weak lensing constraints with stellar kinematics to show that the initial mass function of massive (lens) galaxies is not consistent with light ones such as those advocated for spiral galaxies. (A)

*C. Dark matter substructure*

63. "Evidence for substructure in lens galaxies?" S. Mao and P. Schneider, Mon. Not. Roy. Astron. Soc. **95**, 587-594 (1998). This seminal paper suggested the presence of dark substructure as the cause of the so-called flux-ratio anomalies. (A)
64. "Direct Detection of Cold Dark Matter Substructure," N. Dalal and C.S. Kochanek, Astrophys. J. **572**, 25-33 (2002). In this very influential paper, the authors used a small sample of gravitational lenses observed in the radio to infer limits on the presence of dark substructure. (A)
65. "Detection of a dark substructure through gravitational imaging," S. Vegetti, L.V.E. Koopmans, A.S. Bolton, T.Treu, and R.Gavazzi, Mon. Not. R. Astron. Soc. **408**(4), 1969-1981, (2010). Describes a dark galactic satellite detected at cosmological distances, purely based on the strong gravitational lensing effect. (A)
66. "Effects of dark matter substructures on gravitational lensing: results from the Aquarius simulations," D.D. Xu, et al., Mon. Not. R Astron. Soc. **398**(3), 1235-1253 (2009) . A recent treatment of the substructure problem from a theoretical point of view. The authors compare predictions from numerical simulations with observations of strong lens flux-ratio anomalies and find that strong lenses may even indicate an excess of massive satellites. (A)
67. "The Survival of Dark Matter Halos in the Cluster Cl 0024+16," P. Natarajan, J.-P. Kneib, I. Smail, T. Treu, R.S. Ellis, S. Moran, M. Limousin, and O. Czoske, Astrophys. J. **693**(1), 970-983, (2009). This study of dark matter substructure in a cluster of galaxies, using both strong and weak lensing, covered an unprecedently wide dynamic range of cluster mass density and infalling group environments, and provided comparison with simulations regarding the cluster dark matter density profile. (A)

*D. Cosmic telescopes*

Gravitational lensing magnifies background sources of light. For this reason, mass concentrations like galaxies or clusters of galaxies can be effectively used as cosmic telescope to study fainter and smaller objects that it would be possible in the absence of gravitational lensing. Two examples of this application of gravitational lensing are given below.

68. "A blue ring-like structure, in the center of the A 370 cluster of galaxies," G. Soucail, B. Fort, Y. Mellier, and J.P. Picat, Astron. Astrophys. **172,** L14–L16 (1987). The discovery of the first giant arcs is described in this classic paper; it marks the beginning of the study of cluster strong lenses as cosmic telescopes. (A)
69. "The formation and assembly of a typical star-forming galaxy at redshift z~3," D.P. Stark, A.M. Swinbank, R.S. Ellis, S. Dye, I.R. Smail, and J. Richard, Nature **455**, 775–777 (2008). A very nice recent illustration of the power of gravitational lensing to magnify the distant universe. (A).

*E. Measuring distances in the universe*

Gravitational lensing affects the arrival time of photons from the source to the observer. Using this effect one can measure distances in the universe, as discussed by the following papers.

70. "The Hubble Constant," N. Jackson, Living Rev. Relativ. **10**(4), 52 pages (2007). This is an online review of the expansion of the universe, with emphasis on gravitational lensing-based measurements. (A)
71. "On the possibility of determining Hubble's parameter and the masses of galaxies from the gravitational lens effect," S. Refsdal, Mon. Not. Roy. Astron. Soc. **128,** 307-310 (1964). This classic paper opened up the use of time delays for cosmography. (A)
72. "Dissecting the gravitational lens B1608+656: II. Precision measurements of the Hubble constant, spatial curvature, and the dark energy equation of state," S.H. Suyu, P.J. Marshall, M.W. Auger, S. Hilbert, R.D. Blandford, L.V.E. Koopmans, C.D. Fassnacht, and T. Treu, Astrophys. J. **711** 201–221 (2010). At the time of writing, this is the most accurate measurement of Hubble's constant using a gravitational lens. This rather dense paper explores the most serious sources of systematic error, and includes their mitigation in the final uncertainty estimate. (A)

*F. Finding strong gravitational lenses*

Over the years, many strong lenses have been found serendipitously. However, several targeted lens surveys are worth noting as precursors for the searches being planned. Accessible reviews of some of the methods of strong lens finding can be found in references 53 and 70 (above).
73. "The Cosmic Lens All-Sky Survey - II. Gravitational lens candidate selection and follow-up," I.W.A. Browne, et al., Mon. Not. R. Astron. Soc. **341**, 13-32 (2003). The CLASS survey involved a hierarchical tree of ever-higher resolution radio imaging of flat-spectrum radio sources with unusual morphologies. By focusing on the sources, not the lenses, the team was able to select the first well-defined, statistically complete sample of lensed objects. (A)

74. "The Sloan Lens ACS Survey. I. A Large Spectroscopically Selected Sample of Massive Early-Type Lens Galaxies," A.S. Bolton, S. Burles, L.V.E. Koopmans, T. Treu, and L.A. Moustakas, Astrophys. J. **638,** 703-724, (2006). The most prolific lens search today, the SLACS project involved high-resolution confirmation imaging of a very pure sample of lens candidates, selected for their anomalous SDSS double-galaxy spectra. (A)

75. "The Sloan Digital Sky Survey Quasar Lens Search. I. Candidate Selection Algorithm," M. Oguri, et al., Astron. J. **132**, 999-1013 (2006). The SQLS project was similar to the CLASS survey in that a combination of spectra and imaging was used to make a statistical lens sample, but the data were all taken in the optical and near infrared. (A)

Gravitational lensing on the cluster scale is readily observed in almost all of the most massive systems, provided the image resolution and depth are high enough.

76. "A New Survey for Giant Arcs," J. F. Hennawi, et al., Astronom J. **135**, 664-681 (2008). "Blind" imaging surveys have also been carried out, most notably using SDSS data. Describes the candidate selection and confirmation of 16 new cluster-scale lenses. (A)

Both galaxy-scale and cluster-scale lenses have been found by visual inspection of images. In recent years, the citizen scientists of Galaxy Zoo have been carrying out their own search, re-discovering known lenses and finding interesting new candidates in the SDSS and HST image archives. They can be followed (and joined!) at the Galaxy Zoo forum:

77. Galaxy Zoo Forum, "Possible strong lenses." http://www.galaxyzooforum.org/index.php?topic=6927

*G. Modeling strong lenses*

Researchers in the field use algorithms similar to those in references 42 and 43 above, but using models designed specifically to answer their science questions, and coupled to powerful numerical exploration and optimization engines. While undergraduate exercises are probably best developed with the web-based resources provided above, it may be that a more advanced tools is useful in some situations.

A number of software codes for modeling strong lens systems have been developed, and in some cases made available by their authors. They differ primarily in the type of source they are designed for (extended and point-like), and the mass scale of the lens (galaxy and cluster). Designed for data analysis and inference in galaxy-scale systems, we have, for example:

78. "GravLens," C.R. Keeton. This code provides an extensive range of simply-parameterized mass models, and a high-resolution lens-equation solver for precise point-like image position, flux, and time-delay prediction. Likelihood maximization can be performed. http://redfive.rutgers.edu/~keeton/gravlens (A)

79. "Semilinear Gravitational Lens Inversion," S. J. Warren and S. Dye, Astrophys. J. **590**, 673-682 (2003). While not the first attempt at reconstructing extended lensed sources, this paper provides a good introduction to the topic: much of the galaxy-scale lens modeling work of the last decade builds on this paper. (A)

80. "LENSVIEW: software for modeling resolved gravitational lens images," R. Wayth and R. Webster, Mon. Not. Roy. Astron. Soc. **372**, 1187-1207 (2006). This code finds the best-fitting simple-lens model, while reconstructing an extended source on a pixelated grid, given image data. https://www.cfa.harvard.edu/~rwayth/lensview/Lensview_Home.html (A)

Likewise, for cluster modeling we have:

81. "LensTool," J.-P. Kneib and E. Jullo. The lens and source models are similar to GravLens, but exploration of the parameter space proceeds differently, to cope with the more complex cluster systems. The options are manual iteration (for which a graphical interface is provided), and Markov Chain Monte Carlo sampling. Weak lensing data can also be included in a joint fit. http://www.oamp.fr/cosmology/lenstool (A)
82. "A Bayesian approach to strong lensing modeling of galaxy clusters," E. Jullo, J.-P. Kneib, M. Limousin, A. Eliasdottir, P. J. Marshall, and T. Verdugo, New. J. Phys. **9,** 447, 35 pages (2007). LensTool has been used extensively throughout the literature; this paper provides a good introduction and an overview of its more recently added capabilities. (A)

Purely for theoretical investigations, with no data-fitting capability, is:

83. "GlamRoc," E. A. Baltz. This adaptive mesh ray-tracing code was written to allow exploration of lensing effects near critical points in the source plane, where very high magnifications can be achieved. http://kipac.stanford.edu/collab/research/lensing/glamroc (A)

All of the above software packages require some experience in compiling, linking, and running code written in c on UNIX systems; some standard astronomy subroutine libraries are also required.

## VI. ADVANCED MATERIAL: WEAK LENSING

*A. General weak lensing material*

The following two review articles summarize the primary results in constraining cosmological models using weak lensing data circa 2007. The first article provides a good qualitative summary of the methodology and results, and is a good resource for a student wanting to learn about what weak lensing can tell us about cosmology without having to understand all of the math behind the models, while the second article fills in much of the needed math.

84. "Weak gravitational lensing and its cosmological applications," H. Hoekstra and B. Jain, Annu. Rev. Nucl. Part. Sci. **58**, 99–123 (2008). (I, A)
85. "Cosmology with weak lensing surveys," D. Munshi, P. Valageas, L. van Warebeke, and A. Heavens, Phys. Rep. **462** (3), 67-121 (2008). (A)
86. "Weak gravitational lensing," M. Bartelmann, and P. Schneider, Physics Reports **340** (4-5), 291-472 (2001). This review contains an extremely in-depth discussion of the basics of weak gravitational lensing, with full mathematical derivations of all of the concepts known as of 2000. This review allows us to limit the number of introductory papers in the following sections and concentrate on papers giving more recent developments in the field. (A)
87. "Probing the universe with weak lensing," Y. Mellier, Ann. Rev. Astron. Astrophys. **37**, 127–189 (1999). This review summarizes the first decade of development of weak lensing, tying together the mathematical basis for how lensing works with qualitative descriptions of early results from the field. While some of the specific details of how various aspects of the weak-lensing signal are measured and modeled have changed in the intervening 13 years, this article is still an excellent starting point for someone wanting to the learn the basics of how weak lensing works in practice. (I, A)

*B. Measuring the mass distribution of a cluster of galaxies*

The first use of weak lensing was to measure the mass-density profiles of individual clusters of galaxies: only around a massive cluster is the weak-lensing signal large enough, over a sufficiently wide area, to be detectable above the noise owing to the intrinsic shapes of the many galaxies being lensed. The first set of references below describes various methods used to measure the mass distributions in individual clusters; the second set explores the potentially large errors that can be made when converting the one and two-dimensional mass measurements to a three-dimensional mass. Both can be used to supplement the information in the Bartelmann and Schneider review (ref. 86).

The following three papers present three different methods of doing mass reconstructions based on maximum likelihood or entropy approaches; such methods have the advantage of being able to incorporate additional information beyond just the weak lensing into the reconstruction.

88. "Maximum-entropy weak lens reconstruction: improved methods and application to data," P. J. Marshall, M. P. Hobson, S. F. Gull, and S. L. Bridle, Mon. Not. Roy. Astron. Soc. **335** (4), 1037-1048 (2002). (A)
89. "Strong and weak lensing united," M. Bradac, P. Schneider, M. Lombardi, T. Erben, Astron. Astrophys. **437**, 39-48 (2005). (A)
90. "Reconstruction of Cluster Masses Using Particle based Lensing. I. Application to Weak Lensing," S. Deb, D. M. Goldberg, V. J. Ramdass, Astrophys. J. **687**, 39-49 (2008). (A)

The following papers investigate an error that pervades mass measurements with weak lensing, namely the assumption of a spherical mass model when fitting a non-spherical system. The first two papers give theoretical expectations of the effect from N-body simulations and analytic tri-axial mass models, while the third paper explores whether the departure from the expected concentrations of certain clusters is a result of this effect.

91. "Effects of asphericity and substructure on the determination of cluster mass with weak gravitational lensing,'' D. Clowe, G. De Lucia, and L. King, Mon. Not. Roy. Astron. Soc. **350** (3), 1038-1048 (2004). (A)
92. "A statistical study of weak lensing by triaxial dark matter haloes: consequences for parameter estimation," V. L. Corless, and L. J. King, Mon. Not. Roy. Astron. Soc. **380**, 149-161 (2007). (A)
93. "Subaru Weak Lensing Measurements of Four Strong Lensing Clusters: Are Lensing Clusters Overconcentrated?" M. Oguri, et al., Astrophys. J. **699** (2), 1038-1052 (2009). (A)

The following two papers describe mass measurements of a particular merging cluster system, the "Bullet Cluster," showing how gravitational lensing can be used to measure the mass of a structure that is not in dynamical equilibrium, and without assuming that "mass follows light."

(In this case, it does not!) They also demonstrate how noise in the weak-lensing data can affect the centers of the inferred two-dimensional mass distributions.

94. "A Direct Empirical Proof of the Existence of Dark Matter," D. Clowe, et al., Astrophys. J. **648** (2), L109-L113 (2006). (A)
95. "Strong and Weak Lensing United. III. Measuring the Mass Distribution of the Merging Galaxy Cluster 1ES 0657-558," M. Bradac, et al., Astrophys. J. **652** (2), 937-947 (2006). (A)

The following three papers compare mass measurements from weak lensing and X-ray observations for large samples of clusters. LoCuSS is a large sample of nearby galaxy clusters with good weak lensing and X-ray measurements; Zhang, et al., use these clusters to investigate the agreement between mass estimates based on X-ray and weak lensing, and provide a good discussion of potential sources of errors in both of these methods. The first two papers also provide a good discussion of the differences among various methods of measuring the cluster mass just with weak-lensing data.

96. "A comparison of weak-lensing masses and X-ray properties of galaxy clusters," H. Hoekstra, Mon. Not. Roy. Astron. Soc. **379**, 317-330 (2007). (A)
97. "The Mass-Lx Relation for Moderate Luminosity X-ray Clusters," H. Hoekstra, M. Donahue, C. J. Conselice, B. R. McNamara, and G. M. Voit, Astrophys. J. **726**, 48, 14 pages (2011). (A)
98. "LoCuSS: comparison of observed X-ray and lensing galaxy cluster scaling relations with simulations," Y.-Y. Zhang, et al., Astron. Astrophys. **482** (2), 451-472 (2008). (A)

The following two papers describe a type of lensing that is between the strong and weak-lensing limits: only single images are formed, but lensing introduces a measurable curvature to the galaxy shape. There is some theoretical promise for this "flexion" improving the combination of strong and weak-lensing data, but current attempts to measure the effect have shown it to be difficult.

99. "Weak gravitational flexion," D. J. Bacon, D. M. Goldberg, B. T. P. Rowe, A. N. Taylor, Mon. Not. R. Astron. Soc. **365** (2), 414-428 (2006). (A)
100. "Weak lensing goes bananas: what flexion really measures," P. Schneider, X. Er, Astron. Astrophys. **485** (2), 363-376 (2008). (A)

*C. Measuring cosmological parameters*

Weak lensing can also be used to measure statistically the distribution of mass in structures that are not dense enough to be detected individually in mass reconstructions. This weak-lensing signal is conventionally known as "cosmic shear," and its measurement is one of the primary methods that has been proposed for accurately measuring cosmological parameters such as the matter and energy content of the universe, the curvature of the universe, and the scale of matter density fluctuations.

From 2000-2002, several competing groups produced the first measurements of cosmic shear from surveys each covering ~10 square degrees. A summary of these results can be found in two review articles:

101. "Cosmological weak lensing," Y. Mellier, and L. van Waerbeke, Class. Quantum Grav. **19** (13), 3505–3515 (2002). (A)
102. "Weak gravitational lensing by large scale structures," A. Refregier, Ann. Rev. Astron. Astrophys. **41**, 645-668 (2003). (A)

More recently, a number of larger-scale, higher-fidelity data sets have been analyzed. The last two of the following five papers describe measurements of cosmic shear from a 2 square degree Hubble Space Telescope survey, combined with photometric redshifts of the lensed galaxies. The addition of the photometric redshifts allowed the authors to measure the evolution of the mass power-spectrum, which provides additional information about the cosmological model beyond what the integrated power-spectrum can provide.

103. "Dark Energy Constraints from the CTIO Lensing Survey," M. Jarvis, B. Jain, G. Bernstein, and D. Dolney, Astrophys. J. **644**, 71-79 (2006). This paper presents results from the first analysis of a data set large enough to place significant constraints on the equation of state of dark energy. It includes a good discussion of how the constraints are measured and how systematic errors in the data can affect the results. (A)
104. "Cosmological constraints from the 100-deg$^2$ weak-lensing survey," J. Benjamin, et al., Mon. Not. Roy. Astron. Soc. **381** (2), 702-712 (2007). This paper presents results from a combination of 4 blank-field surveys to achieve a 100 square degree total area, and contains a good discussion on how errors in Point Spread Function correction and background galaxy selection can propagate through to the cosmic-shear measurement. (A)
105. "Very weak lensing in the CFHTLS wide: cosmology from cosmic shear in the linear regime," L. Fu, et al., Astron. Astrophys. **479**, 9-25 (2008). This paper describes the results of a cosmic-shear measurement from the third-year release of the CFHT Legacy Survey data. It provides a summary of basic measurement methodology, and a good discussion of potential sources of systematic error. (A)

106. "COSMOS: Three-dimensional Weak Lensing and the Growth of Structure," R. Massey, et al., Astrophys. J.S **172**, 239-253 (2007). (A)
107. "Evidence of the accelerated expansion of the Universe from weak lensing tomography with COSMOS," T. Schrabback, Astron. Astrophys. **516**, A63 (2010). (A)

The first of the following three papers describes an attempt to detect clusters by mass in a blank-field survey, thereby avoiding selection bias when measuring the evolution of the mass function. The last two papers describe how to use such techniques to measure the evolution of the power spectrum, and thereby cosmological parameters, in a method complementary to cosmic-shear analysis.

108. "First Results on Shear-selected Clusters from the Deep Lens Survey: Optical Imaging, Spectroscopy, and X-Ray Follow-up," D. Wittman, et al., Astrophys. J. **643**, 128-143 (2006). (A)
109. "Cosmology with the shear-peak statistics," J. P. Dietrich, and J. Hartlap, Mon. Not. R. Astron. Soc. **402** (2), 1049-1058 (2010). (A)
110. "Probing cosmology with weak lensing peak counts," J. M. Kratochvil, Z. Haiman, and M. May, Phys. Rev. D **81** (4), 043519, 16 pages (2010). (A)

The following three papers discuss the impact on cosmological-parameter measurement from cosmic-shear surveys owing to a variety of systematic errors. Of particular importance are the intrinsic alignments of galaxy shapes (prior to their being lensed) and the accuracy with which the lensed-galaxies redshifts can be estimated from their colors alone.

111. "Dark energy constraints from cosmic shear power spectra: impact of intrinsic alignments on photometric redshift requirements," S. Bridle, L. King, New J. Phys. **9** (12), 444-468 (2007). (A)
112. "Systematic effects on dark energy from 3D weak shear," T. D. Kitching, A. N. Taylor, A. F. Heavens, Mon. Not. Roy. Astron. Soc. **389**, 173-190 (2008). (A)
113. "Catastrophic photometric redshift errors: weak-lensing survey requirements," G. Bernstein, D. Huterer, Mon. Not. Roy. Astron. Soc. **401** (2), 1399-1408 (2010). (A)

Recently, a number of papers have explored using weak lensing magnification, rather than shear, as a tool to measure cosmological parameters.

114. "Detection of Cosmic Magnification with the Sloan Digital Sky Survey," R. Scranton, et al., Astrophys. J. 633, 589-602 (2005). This paper studies the change in number density of quasars with distance from bright galaxies and models the results as a gravitational lensing signal from the bright galaxies. It includes a discussion of the potential systematic errors in the measurements and why previous attempts at making this measurement were limited by the systematic errors. (A)
115. "CARS: The CFHTLS-Archive-Research Survey. III. First detection of cosmic magnification in samples of normal high-z galaxies," H. Hildebrandt, L. van Waerbeke, and T. Erben, Astron. Astrophys. 507 (2), 683-691 (2009). This paper studies how the distribution of faint, background galaxies is affected by the presence of bright, foreground galaxies, and models this as a product of weak lensing magnification. (A)
116. "Shear and magnification: cosmic complementarity," L. van Waerbeke, Mon. Not. R. Astron. Soc. 401 (3), 2093-2100 (2010). This paper described the methodology behind using weak lensing magnification of high-redshift galaxies as a probe of comological parameters. (A)

*D. Measuring the average mass distribution in stacked galaxies*

The following papers demonstrate how to use the weak-lensing signal around a "stacked" ensemble of foreground galaxies to measure their mean mass profile out to a much larger radius than can be achieved with strong-lensing measurements.

117. "Weak Gravitational Lensing by Galaxies," T. G. Brainerd, R. D. Blandford, and I. Smail, Astrophys. J. **466**, 623-637 (1996). This paper introduced the concept of galaxy-galaxy lensing, and provides scaling relations for predicting the number of background galaxies that will be needed to measure the mean mass profile around a given type of galaxy to a given accuracy. (A)
118. "Properties of Galaxy Dark Matter Halos from Weak Lensing," H. Hoekstra, H. K. C. Yee, and M. Gladders, Astrophys. J. **606**, 67-77 (2004). A study of galaxy-galaxy lensing using a 45 square degree imaging survey, including the first significant attempt to measure the shape of the dark-matter halo by rotating (as well as translating) the background galaxy positions and shears to a common reference frame. (A)
119. "The Galaxy-Mass Correlation Function Measured from Weak Lensing in the Sloan Digital Sky Survey," E. Sheldon, et al., Astron. J. **127** (5), 5244-2564 (2004). This presents a galaxy-galaxy lensing study using a database large enough to subdivide the foreground galaxy sample by various properties, and still obtain a high signal-to-noise measurement of the mass profiles. (A)
120. "Ellipticity of dark matter haloes with galaxy-galaxy weak lensing," R. Mandelbaum, et al., Mon. Not. Roy. Astron. Soc. **370** (2)**,** 1008-1024 (2006). This paper includes measurements of the ellipticity of dark-matter halos for different foreground galaxy properties. It includes a good discussion of potential systematic errors coming from intrinsic alignments of the background galaxies. (A)
121. "New constraints on the evolution of the stellar-to-dark matter connection: a combined analysis of galaxy-galaxy lensing, clustering, and stellar mass functions from z=0.2 to z=1,'' A. Leauthaud, et al., Astrophys. J. **744** (2), 159, 27 pages (2012). This paper uses a large Hubble Space Telescope survey to obtain galaxy-galaxy lensing measurements at higher redshifts than previously possible, and explores the evolution of dark halo properties. (A)

*E. Technical challenges*

All of the above studies rely upon our ability to detect faint galaxies, measure their shapes, and correct those shape measurements for the effects of smearing by the atmosphere and the telescope optics. The point-spread-function (PSF) correction is usually the largest source of systematic error in modern weak-lensing measurements, and much of the effort of the community is devoted to mitigating it. Instead of citing individual papers describing the various methods, we instead list below the reports from large collaborations carrying out blind analysis of simulated data to test the various methods. These papers provide good summaries of these methods, and contain references that can be followed to learn more about any specific method.

122. "The Shear Testing Programme - I. Weak lensing analysis of simulated ground-based observations," C. Heymans, et al., Mon. Not. Roy. Astron. Soc. **368** (3), 1323-1339 (2006). (A)
123. "The Shear Testing Programme 2: Factors affecting high-precision weak-lensing analyses," R. Massey, et al., Mon. Not. Roy. Astron. Soc. **376**, 13-38 (2007). (A)
124. "Results of the GREAT08 Challenge: an image analysis competition for cosmological lensing," S. Bridle, et al., Mon. Not. Roy. Astron. Soc. **405** (3), 2044-2061 (2010). (A)

Information about subsequent simulations that had not produced final reports at the time of writing this Resource Letter can be found at http://greatchallenges.info/.

*F. Publicly available software*

Software for measuring galaxy shapes in images, including the PSF correction, are listed below. All come with manuals or help files to some degree, but all also have severe learning curves if they are to be used effectively. We strongly encourage testing the software on the simulations described in the section above before using them on actual data.

125. Imcat http://www.ifa.hawaii.edu/~kaiser/imcat
126. KSBf90 http://www.roe.ac.uk/~heymans/KSBf90
127. im2shape http://www.sarahbridle.net/im2shape
128. lensfit http://www.physics.ox.ac.uk/lensfit
129. shapelets http://www.astro.caltech.edu/~rjm/shapelets

Software for performing one and two-dimensional mass reconstructions in individual fields are listed below.

130. Imcat http://www.ifa.hawaii.edu/~kaiser/imcat
131. LensEnt2 http://www.slac.stanford.edu/~pjm/lensent/version2
132. Particle-Based Lensing http://www.physics.drexel.edu/~deb/PBL.htm
133. Lenstool http://www.oamp.fr/cosmology/lenstool (One needs to turn off the strong lensing portion to perform just a weak lensing measurement.)
134. MRLENS http://irfu.cea.fr/Sap/en/Phocea/Vie_des_labos/Ast/ast_visu.php?id_ast=878

There is no publicly available package to fully perform a cosmic-shear analysis of a data set. However, some very useful tools can be found at Martin Kilbinger's webpage under his software section: http://www2.iap.fr/users/kilbinge

**VII. ADVANCED MATERIAL: MICROLENSING**

While the study of strong lensing can be traced back to Zwicky's 1937 paper (ref. 27), microlensing is a field that was perhaps opened up more directly by Einstein and Eddington's initial work. Indeed, microlensing is presumed to be, and in many cases observed to be, due to stars. After providing some references to some general material on the topic, we organize the rest of this section by the location of these stars - either in our own galaxy, or in external galaxies that are themselves strong lenses.

*A. General microlensing*

135. "Gravitational Microlensing in the Local Group", B.Paczynski, Ann. Rev. Astron. Astrophys. **34**, 419-459 (1996). A comprehensive review of microlensing, covering the history of the field, its theoretical foundations, a critical review of observational results at the time of its writing and a discussion of future prospects. (A)

*B. Galactic microlensing*

The idea that gravitational lensing magnification of background stars by massive compact objects in the Milky Way halo generated a lot of interest in the 1980s and 1990s. The search for this phenomenon required multi-epoch imaging surveys; when they came, they provided the best upper limits we have on dark matter taking this particular form. An excellent introduction and review is given by reference 44. Also worthy of note is the following paper.

> 136. "Gravitational microlensing by the galactic halo," B. Paczynski, Astrophys. J. **304**, 1–5 (1986). The foundational paper of galactic microlensing. The basic features of galactic microlensing and its potential as a probe of massive compact objects as dark matter are discussed. This influential paper spurred a number of gravitational-lensing experiments, such as the one described in the two following papers. (A)

Whereas stars are the main deflectors found by microlensing searches, a small fraction of them turns out to have planets in orbit around them, as evidenced by the short timescale perturbations to the expected magnification variations. This technique for planet detection continues to attract attention, especially since it is one of those that have the potential to be sensitive to Earth-like planets in the habitable zone. The following two references describe planet microlensing.

> 137. "Discovery of a cool planet of 5.5 Earth masses through gravitational microlensing," J.-P. Beaulieu, et al., Nature **439** (7075), 437–440 (2006). The fascinating discovery of an Earth-like planet using gravitational microlensing. (A)
> 138. "Studying planet populations with Einstein's blip," M. Domink, Phil. Trans. R. Soc. A. 368, 3535-3550 (2010). A recent review specialized in microlensing as a tool to study planets. (A).

*C. Cosmological microlensing*

Strongly-lensed quasars have been observed to show additional, independent variability in their four images; this can often be attributed to microlensing by stars in the lens, and successfully modeled as such. Indeed, this effect, and in particular its variation with wavelength, has provided an opportunity to study the source quasars' central engines in unprecedented detail, while the magnitude of the microlensing variability has been used to constrain the stellar density in the lens. An excellent introduction is given by reference 44. Here we report two recent articles to describe the state of the art of the field.

> 139. "X-Ray And Optical Flux Ratio Anomalies In Quadruply Lensed Quasars. II. Mapping the Dark Matter Content in Elliptical Galaxies" D.Pooley, S.Rappaport, J. Blackburne, and J.Wambsganss, Astrophys. J. **744** (2), 111, 13 pages (2012). Used microlensing to learn about the granularity of the mass distribution in galaxies and therefore infer an absolute measurement of the dark matter and stellar content. This is a beautiful example of single-epoch microlensing applications. (A)
> 140. "Quantitative interpretation of quasar microlensing light-curves", C.S. Kochanek, Astrophys. J. **605**, 58-77 (2004). Introduced a new method to analyze and interpret quasar microlensing light curves. In the process it gave a good description of the physical picture. (A)

**Acknowledgments**

Much of this Resource Letter was planned during the 2011 Summer School on Gravitational Lensing held at National Astronomical Observatory China, Beijing, China. We are very grateful to Professor Hu Zhan and his colleagues for organizing the school, and for their gracious hospitality. We are also extremely grateful for our fellow lecturers, M. Bartelmann, G. Bernstein, A. Heavens, V. Margonniner, M. Takada, and D. Wittmann, for the key input they provided. We also thank R. Massey and C. Heymans for their help in identifying some of the available software resources and providing useful feedback on the sections about weak lensing. We thank Sherry Suyu for comments on an early version of this manuscript. We are also grateful to the Resource Letters editor Dr R.H.Stuewer, and the reviewers for feedback and comments that improved this manuscript. One of us (T.T.) acknowledges support by the Packard Foundation through a Packard Research Fellowship, and one of us (P.J.M.) acknowledges support from the Royal Society in the form of a research fellowship.

21